\documentclass[a4paper,11pt]{article}
\usepackage{jinstpub} 
\usepackage{lineno}
\usepackage{placeins}
\usepackage{subcaption}

\title{Studies of ultrafast dynamics in substrate-free nanoparticles at ELI using Timepix3 optical camera}

\author[a]{Dmitrij Ševaev,}
\author[a,b,c]{Andrei Nomerotski,}
\author[a,b]{Peter Švihra,}
\author[d,e]{Keshav Sishodia,}
\author[d]{Andreas~Hult Roos,}
\author[d]{Martin~Albrecht,}
\author[e]{Sivarama Krishnan,}
\author[d]{Jakob~Andreasson,}
\author[d,f]{Maria~Krikunova,}
\author[d]{Eva~Klimešová}

\affiliation[a]{Faculty of Nuclear Sciences and Physical Engineering, Czech Technical University in Prague,
	Břehová 7, Prague, Czech Republic}
\affiliation[b]{Institute of Physics of the Czech Academy of Sciences,
	Na Slovance 1999/2, Prague, Czech Republic}
\affiliation[c]{Department of Electrical and Computer Engineering, Florida International University,
	10555 West Flagler St, Miami, U.S.A}
\affiliation[d]{ELI Beamlines facility, The Extreme Light Infrastructure ERIC, Za Radnicí 835, 252 41 Dolní Břežany, Czechia}
\affiliation[e]{Quantum Center of Excellence for Diamond and Emergent Materials, School of Interdisciplinary Sciences and Department of Physics, Indian Institute of Technology Madras, 600036 Chennai, India}
\affiliation[f]{Technical University of Applied Sciences, Hochschulring 1, 15745 Wildau, Germany}

\emailAdd{andrei.nomerotski@cvut.cz}

\abstract{We present a novel application of the Timepix3 optical camera (Tpx3Cam) for investigating ultrafast dynamics in substrate-free nanoparticles at the Extreme Light Infrastructure European Research Infrastructure Consortium (ELI ERIC). The camera, integrated into an ion imaging system based on a micro-channel plate (MCP) and a fast P47 scintillator with nanosecond scale emission time, enables individual time-stamping of incoming ions with nanosecond timing precision and high spatial resolution.
The detector successfully captured laser-induced ion events originating from free nanoparticles disintegrated by intense laser pulses. Owing to the broad size distribution of the nanoparticles (10–500 nm) and the variation in laser intensities within the interaction volume, the detected events range in occupancy from near-zero to extremely high—approaching the readout limits of the detector.
By combining time-of-flight and velocity map imaging (VMI) techniques, detailed post-processing and analysis were performed. The results presented here focus on the performance of Tpx3Cam under high-occupancy conditions, which are of particular relevance to this study. These conditions approach the limitations imposed by the camera’s readout capabilities and challenge the effectiveness of standard post-processing algorithms.
We investigated these limitations and associated trade-offs, and we present improved methods and algorithms designed to extract the most informative features from the data.}


\begin{document}
\maketitle
\flushbottom

\section{Introduction}
\label{sec:introduction}
Velocity map imaging (VMI) is a well-established technique for measuring momenta of electrons and ions in gas phase processes \cite{eppink1997,mabbs2008,vallance2019}. VMI is widely used, e.g. in physical chemistry to obtain photoelectron distributions in the molecular frame \cite{reid2012} and follow chemical dynamics \cite{vallance2019} or for temporal diagnostics of x-ray pulses from a free electron laser \cite{duris2020}. In a VMI spectrometer, a set of electrodes is used to create an electric field that guides the charged fragments (electrons or ions) to a position-sensitive detector, where the velocity of the charged particles is mapped onto the position on the detector. The detector typically consists of a microchannel plate (MCP) with a phosphor screen, which is imaged on an optical camera. The optical camera (commonly a slow CMOS camera) captures the image of the phosphor but does not provide any temporal information.

To improve the VMI setup, the standard CMOS camera can be replaced with a time-stamping pixel detector, such as Timepix3 data-driven optical camera \cite{timepixcam, zhao2017, nomerotski2019}. This allows for time stamping of each detector hit and thus obtaining spatial and temporal distribution of the detected ions. Simultaneous spatial and temporal data acquistion at the VMI spectrometer opens novel detection schemes, such as multi-mass VMI \cite{vallance2019}, covariance-map imaging \cite{lee2020,allum2021,schouder2021} and 3D imaging of ions or electrons \cite{basnayake2022,cheng2022,fisher-levine2018}.

\begin{figure}[htbp]
\centering
\hfill
\includegraphics[width=\textwidth]{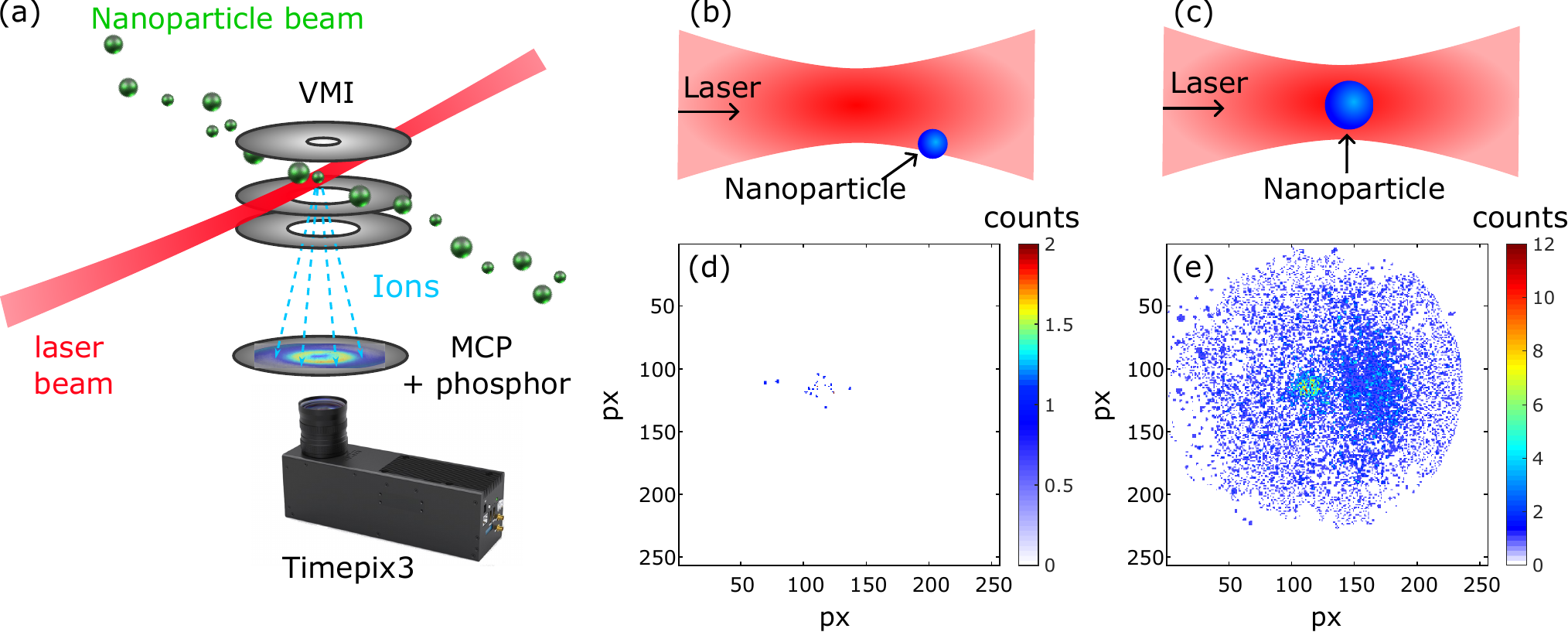}
\caption{
(a) Experimental setup: schematic representation of the laser beam interacting with a nanoparticle beam in the gas phase \cite{Klimesova2025slides}. Nanoparticles are much smaller than the laser focal spot. Therefore, each nanoparticle is irradiated with a different laser intensity. Produced ions are projected to the MCP \& phosphor (P47) ion detector and are imaged with the Timepix3 camera. (b, c) Examples of positions of nanoparticles (blue) within the laser focal volume (red) during the nanoparticle-laser interaction. (b) A small nanoparticle is hit in a region of lower laser intensity, producing an event with low occupancy. (c) A large nanoparticle is hit in a region of high laser intensity, producing an event with very high occupancy. (d) An example of recorded single-shot image with low occupancy, corresponding to the situation in (b). Individual ion hits are well resolved. (e) Example of single-shot image with a very high occupancy, corresponding to the situation in (c). Blending of individual ion hits is clearly visible.}
\label{fig:sketch}
\end{figure}

In this work, we critically assess and analyse the performance of an optical Timepix3 camera at VMI spectrometer for an ion measurement, where occupancies range from almost zero to extremely high (up to situations with all pixels hit in an event). In the experiment performed at the ELI Beamlines facility (ELI ERIC), nanoparticles were irradiated with a strong laser beam to produce ions with different spatial distributions. The aim of the experiment was to investigate mechanisms of ion emission at the nanoscale. In the interaction of the laser beam with the nanoparticle beam, generally, each laser shot interacts with a nanoparticle of different size at a different position in the focal volume (corresponding to a different local laser intensity). Random nature of the interaction results in a strongly varying signal levels, see Figure \ref{fig:sketch} for a schematic diagram and examples of events with largely different occupancies. These signal fluctuations are common in gas phase experiments with nanoscale systems \cite{klimesova2025,medina2023} and therefore it is of high importance to develop data acquisition systems and data analysis protocols that can efficiently handle large variations of signal levels.

Our analysis of images with high occupancies shows that the readout capabilities of the detector were reaching their limit. In the following we present advanced analysis of experimental data and extraction of critical features from the data. Overall we demonstrate that an overflow of the signal for the highest occupancy shots can be successfully corrected.

\section{Methods}
\label{sec:methods}
\subsection{Experiment with krypton nanoparticles}
The experiment was performed at the MAC end-station at the ELI Beamlines facility (ELI ERIC) \cite{klimesova2021mac,klimesova2024}, using a Ti:Sapphire laser system (Coherent, Legend Elite Duo). For the presented experiment, we used laser pulses at the central wavelength of 800~nm, pulse energy $\approx 100\;\mu$J, pulse duration of 40~fs at a repetition rate of 1~kHz. The laser beam was focused to the peak intensity of $1\times 10^{14}$\;W\,cm$^{-2}$. A beam of krypton (Kr) nanoparticles was produced by Even-Lavie pulsed valve \cite{even2015} cooled down to 200~K at the Kr backing pressure of 50~bar. The nanoparticle beam was skimmed twice to produce a low-density stream of nanoparticles in the interaction region. On average, less than one nanoparticle was intercepted by each laser shot. Ions created in the interaction of the laser beam with nanoparticles were detected with a VMI spectrometer equipped with an ion detector and Timepix3 optical camera (Tpx3Cam, Amsterdam Scientific Instruments, see Figure~\ref{fig:sketch}(a) and Figure~\ref{fig:diagram}).

In the VMI spectrometer, different ion species are separated in time according to their mass to charge ratio: lighter ions arrive earlier at the detector and heavier ions later. From the temporal arrival of the ions, we extracted their time of flight (ToF), see Section~\ref{subsec:data_analysis}, to obtain the ToF mass spectrum. Different ion species form distinct peaks in the ToF spectrum. The spatial distributions of selected ions provide information about angular distributions of their velocities.

The Even-Lavie valve operated at a repetition rate of 500~Hz, synchronized with the laser beam running at 1~kHz. The temporal width of the nanoparticle pulse was longer than the 1~ms period of the laser beam due to the broad size distribution of krypton nanoparticles. This setting resulted in a production of two interaction sequences, each at 500~Hz: a "Large sequence", where a laser pulse interacted with larger nanoparticles in the beam and a second "Small sequence", where the laser pulse interacted with smaller nanoparticles within the same nanoparticle pulse. In this work, we focussed on the analysis of the large sequence as it contained high occupancy events.

\subsection{Ion detection with Timepix3 optical camera}
\label{subsec:camera}

\begin{figure}[htbp]
\centering
\includegraphics[width=1.\textwidth]{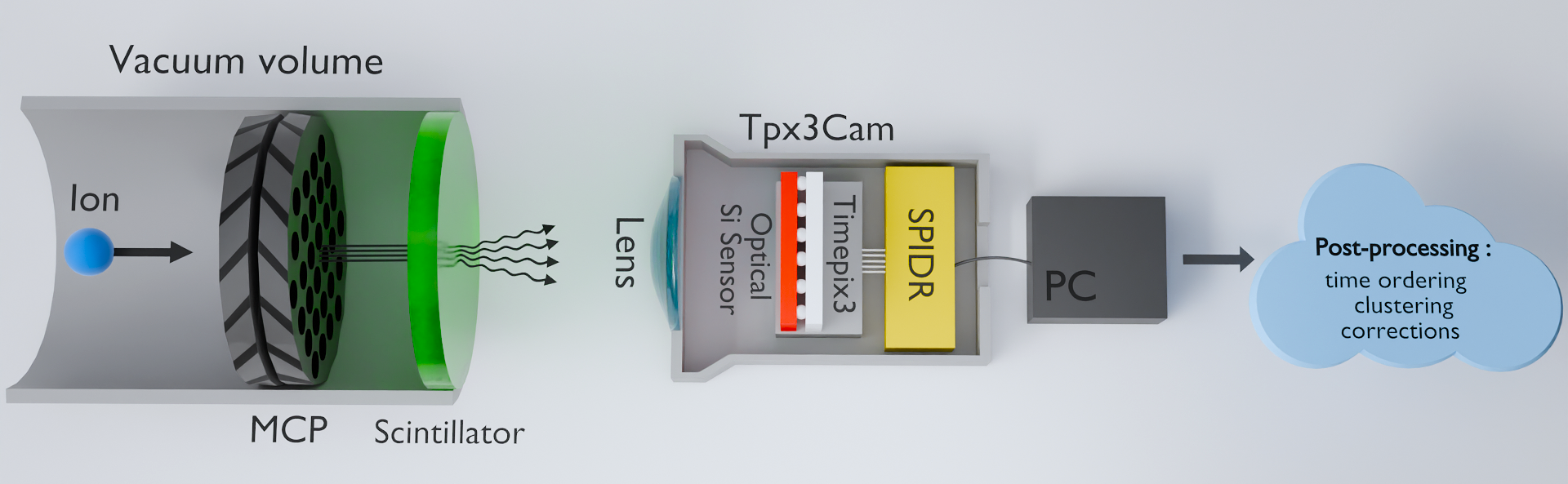}
\caption{Schematic diagram of the ion detector illustrating the use of the Timepix3 camera in conjunction with the velocity map imaging (VMI) technique. \label{fig:diagram}}
\end{figure}

Figure \ref{fig:diagram} shows a schematic layout of the used ion detector with a microchannel plate (MCP) followed after a gap by a thin layer of P47, a fast scintillator with 7~ns risetime \cite{winter2014}. The ion signal was amplified by the MCP and converted to a fast light flash by the  scintillator. The experiments were performed using a camera based on the Timepix3 hybrid pixel detector, Tpx3Cam \cite{timepixcam, zhao2017, nomerotski2019}, where a specialized optical sensor with high quantum efficiency \cite{Nomerotski2017} is attached to the Timepix3 readout chip \cite{Poikela2014}. The chip consists of a 256 × 256 array of 55 µm pixels, each with its own charge amplifier, discriminator, and digital logic to record time of arrival (ToA) and time over threshold (ToT) per hit with timing binning of 1.56 ns.

In contrast to frame-based detectors, Timepix3 implements a fully data-driven sparse readout: only pixels that detect a signal are transmitted off-chip. This architecture allows for high sustained hit rates without compromising time resolution or multi-hit capability.
The on-chip output is organized into eight parallel serial data links at 160 Mb/s each and uses the SPIDR (Speedy PIxel Detector Readout) interface to output the data \cite{heijdenSPIDR}. This readout system supports an effective throughput of up to 80 Mpixel/s with 10~Gbs optical link. In these measurements the user-level rate was sustained at about 40~Mpixel/s.

\subsection{Data analysis pipeline}
\label{subsec:data_analysis}

The measured data were transferred from the Timepix3 camera to a PC and subsequently post-processed using the following steps.
\begin{itemize}
  \item Raw hit data from the Timepix3 camera are time-ordered, and then clustered into spatially and temporally contiguous pixel groups corresponding to detected ions \cite{zhao2017}.
  \item Centroiding is applied to each cluster to extract the effective hit position and associated uncertainties \cite{zhao2017}.
  \item Each centroided cluster is characterized by 15 parameters, including trigger information, pixel coordinates, time of arrival (ToA), time over threshold (ToT), cluster size, centroid position $(x,y)$ and their respective standard deviations.
  \item Cluster ToA and ToT are assigned from the pixel with the highest ToT, and the time of flight (ToF) is computed relative to the laser synchronization pulse.
  \item The resulting ToF values are used to determine the ion mass to charge ratio.
\end{itemize}

Data analysis was performed in the Python programming environment using its standard scientific libraries. This workflow enables both a global overview of the measured data and highly selective filtering algorithms.

\section{Results}
\label{sec:results}

The results presented in this section can be divided into two categories:
(a) performance of the camera, including its capabilities, limitations, and the applied corrections; and
(b) scientific analysis, illustrating the broad range of data exploration possibilities enabled by the recorded datasets.
The primary focus of this study was imaging ultrafast ion dynamics, requiring detection and identification of individual ions.

\subsection{Ion mass spectrum}

\begin{figure}[htbp]
\centering
\includegraphics[width=.85\textwidth]{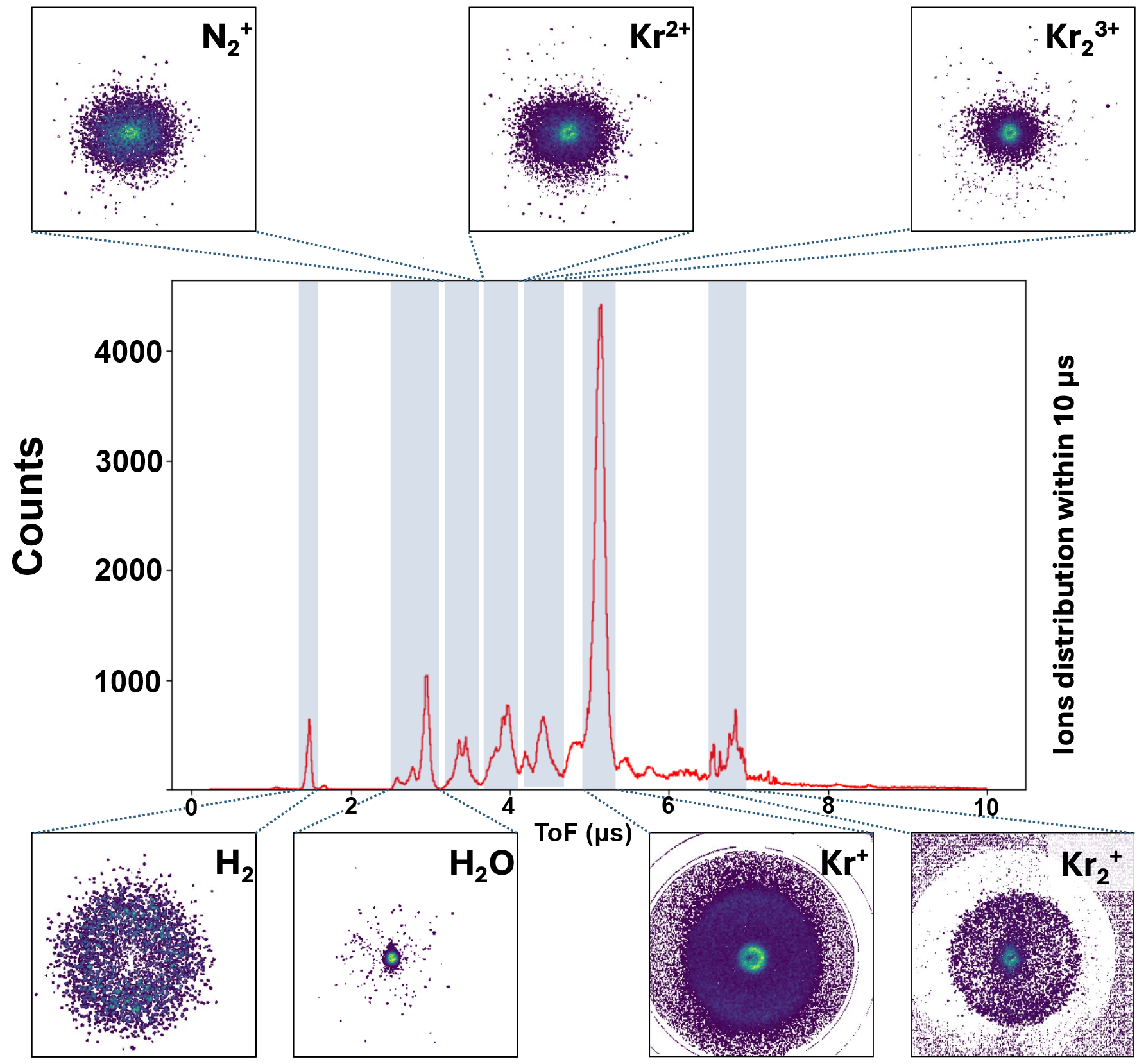}
\caption{Time of flight (ToF) mass spectrum of ions detected within the first 10 µs after the laser–nanoparticle interaction. The red histogram shows the ion count distribution as a function of ToF. Distinct peaks are observed, each corresponding to different ion species separated according to their mass to charge ratio. The shaded regions highlight characteristic ToF windows associated with selected ion species. Zoomed images display integrated ion hit patterns on the detector for selected ToF regions, including light fragments from the background gas as well as ions from krypton nanoparticles. \label{fig:ions}}
\end{figure}

Figure \ref{fig:ions} shows the full time-of-flight spectrum of measured ions.
Distinct peaks are observed, each corresponding to different ion species separated according to their mass to charge ratio. The shaded regions highlight characteristic ToF windows associated with selected ion species. Zoomed images display integrated ion (x,y) hit patterns on the detector for selected regions, including light molecular fragments from the background gas as well as ions from krypton nanoparticles. Such approach allows us to study the overall experimental trends, detailed behaviour of individual ions and correlations between different ion species as ions of all masses are acquired simultaneously.

The ion images reveal the typical spatial distribution of the detected events, reflecting both the Timepix3 detection capabilities and the kinematics of the ion emission. The main peak of Kr$^+$ ions is clearly visible at 5.2 µs, which was a particular focus of this study providing high occupancy events.

\FloatBarrier
\FloatBarrier

\subsection{Correction of shifted time-stamps}

\begin{figure}[htbp]
    \centering
    \includegraphics[width=\textwidth]{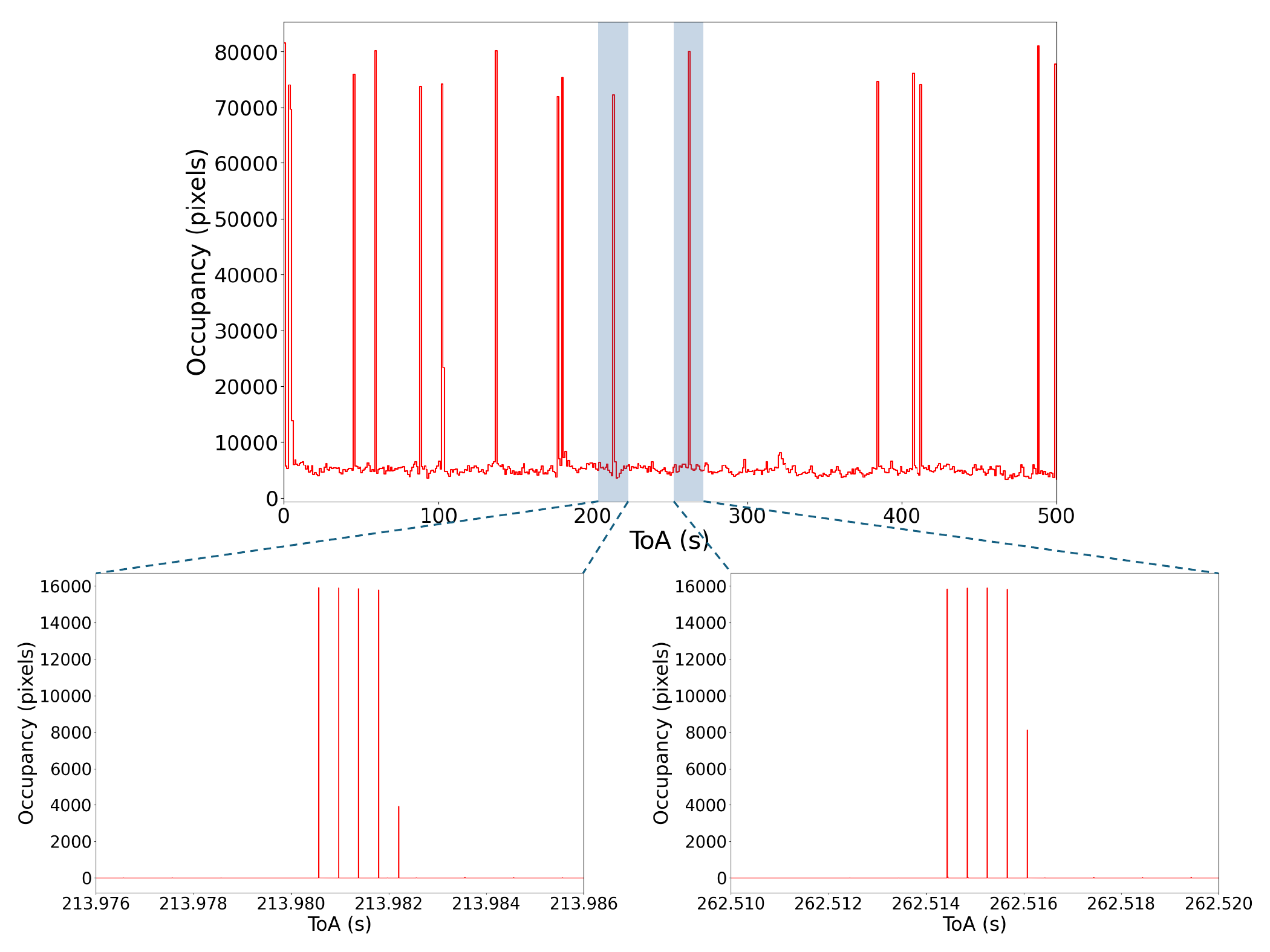}
    \caption{Histograms of hit occupancies for large sequences. The upper panel displays the overall distribution of hits recorded during the 500 s acquisition. A closer inspection of the peak structure reveals that each peak is composed of several sub-peaks, originating from the 409.6 µs timestamp shift. These sub-peaks correspond to the same physical event, registered with timestamps displaced by integer multiples of 409.6 µs. The algorithm introduced in the text successfully identifies and corrects these shifted contributions, yielding the results presented in Figure {\ref{fig:comparison}}.
}
    \label{fig:rates}
\end{figure}

During the analysis, at the highest occupancies a readout saturation was observed: distinct peaks appeared in the ToF spectra at positions shifted from their expected locations by exact multiples of 409.6 µs. This value is equal to 1.5625~ns $\times ~2^{18}$ corresponding to the maximum ToA value, which can be measured by Timepix3 itself. After this limit is reached, ToA needs to be extended by the SPIDR readout system, outside of Timepix3. If the instantaneous rate is very large and the readout takes more time than this value, then the shift can occur in the used version of readout system.
A detailed examination of both nominal and shifted peaks indicated that the latter originate from standard peaks of the acquisition sequences. The shifted peaks were displaced by 409.6 µs due to the large rate and subsequent readout overflow as illustrated in Figure \ref{fig:rates}. After all hits are ingested, the readout performance returns to the usual operation.

\FloatBarrier
\FloatBarrier

\begin{figure}[htbp]
    \centering
    \includegraphics[width=0.7\textwidth]{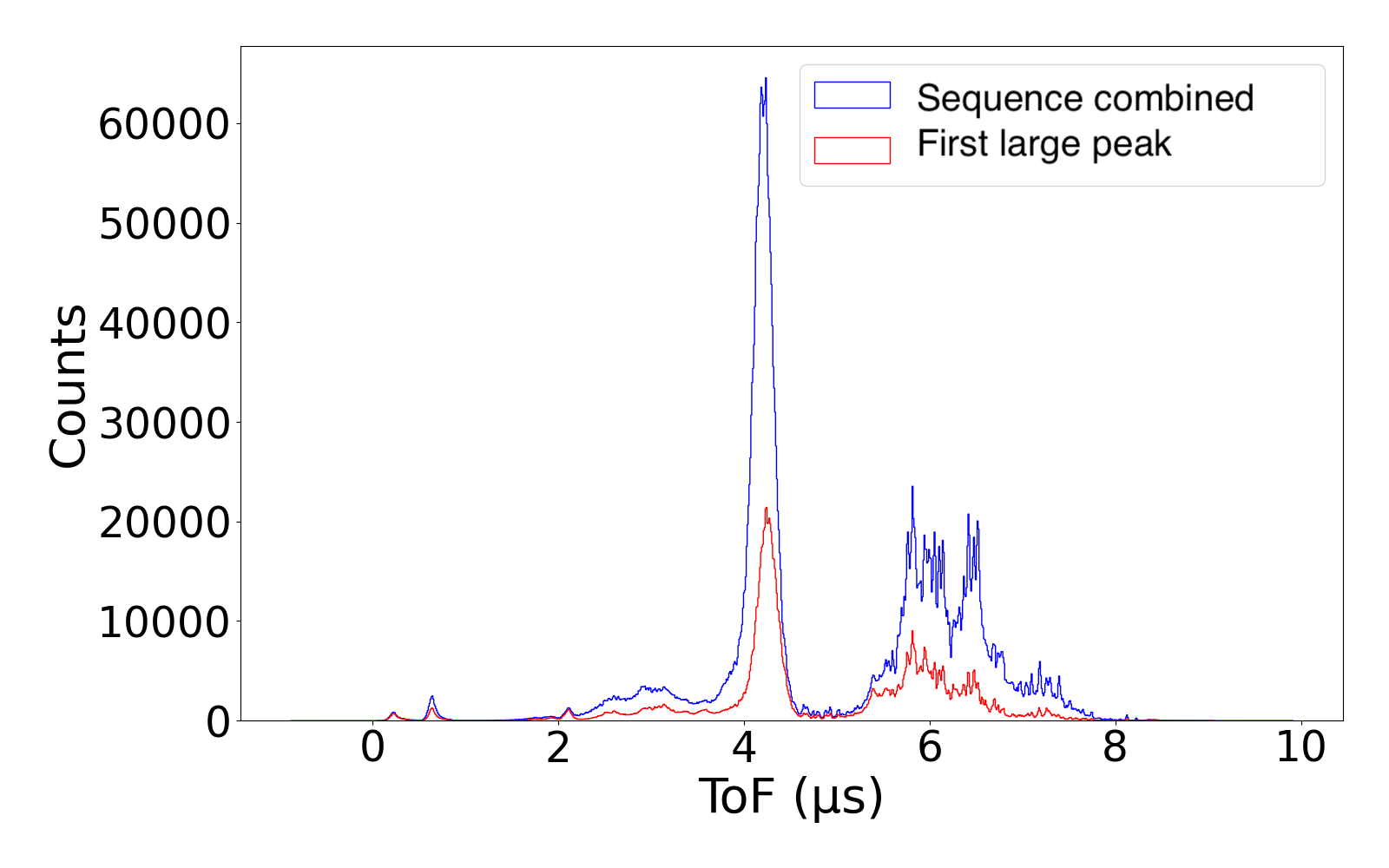}
    \caption{Histogram of the corrected peak for the large sequence. The correction algorithm identifies peaks shifted by 409.6 µs and aligns them by subtracting this offset. The resulting histogram is then shifted back by the laser period (1.000935 ms) to ensure proper temporal alignment, illustrating the effectiveness of the correction procedure.}
    \label{fig:comparison}
\end{figure}

To correct for this effect, a dedicated algorithm was developed. It identifies the shifted peaks, associates them with their original sequence, then subtracts the determined shift from their time-stamps and merges them into a single reconstructed peak.
The effectiveness of this correction procedure is illustrated in Figure \ref{fig:comparison}, where the first peak and its corrected sequence are overlaid, demonstrating good agreement of the resulting ToF distributions. This tool enables reconstruction of data affected by Timepix3 readout overflow and allows subsequent analysis to proceed on corrected datasets. The overflow mechanism has been investigated and will be corrected in future updates of the readout system.

\FloatBarrier
\FloatBarrier

\subsection{Event multiplicity}
\begin{figure}[htbp]
\centering
\begin{subfigure}[b]{0.32\textwidth}
    \includegraphics[width=\textwidth]{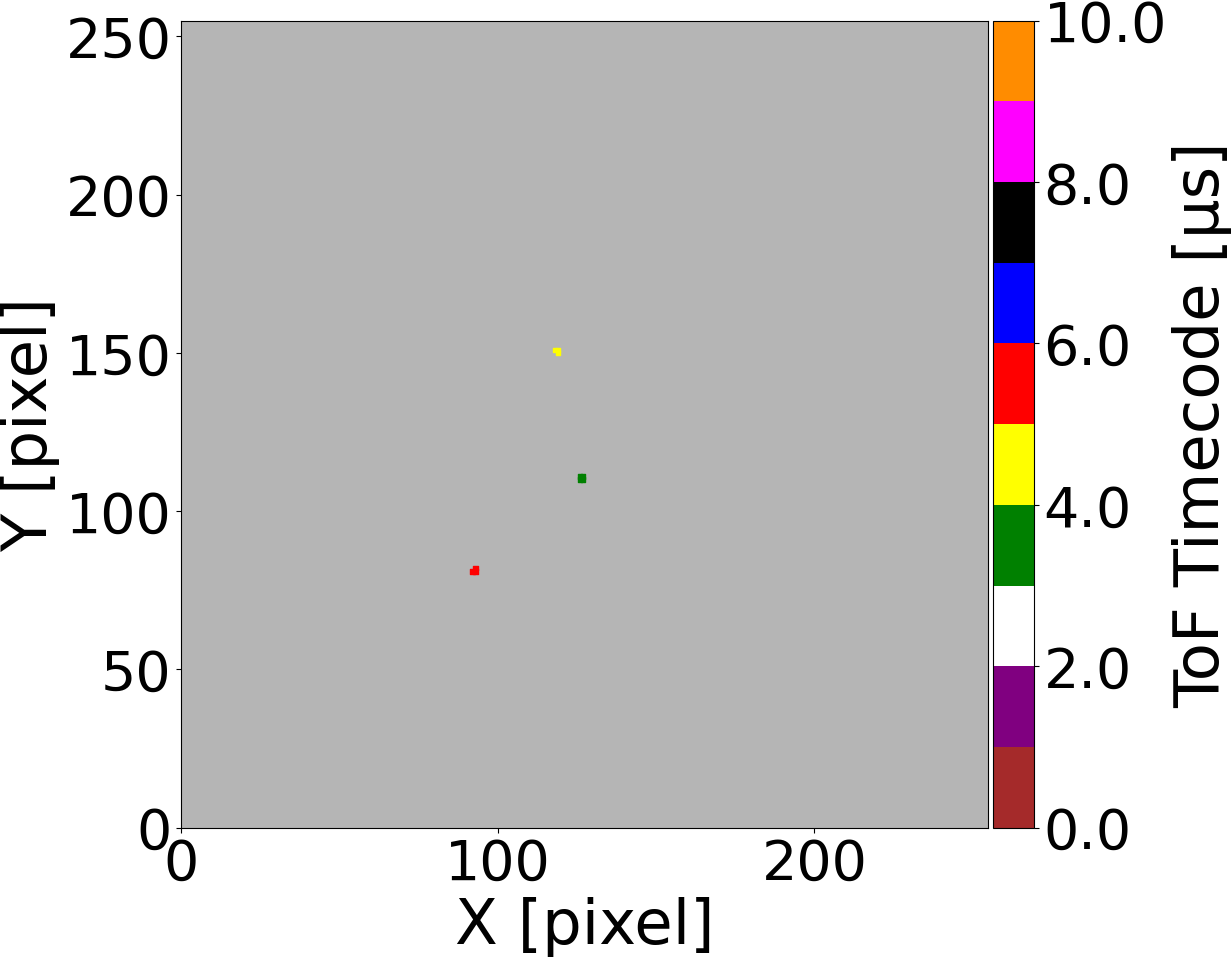}
\end{subfigure}\hspace{1mm}
\begin{subfigure}[b]{0.32\textwidth}
    \includegraphics[width=\textwidth]{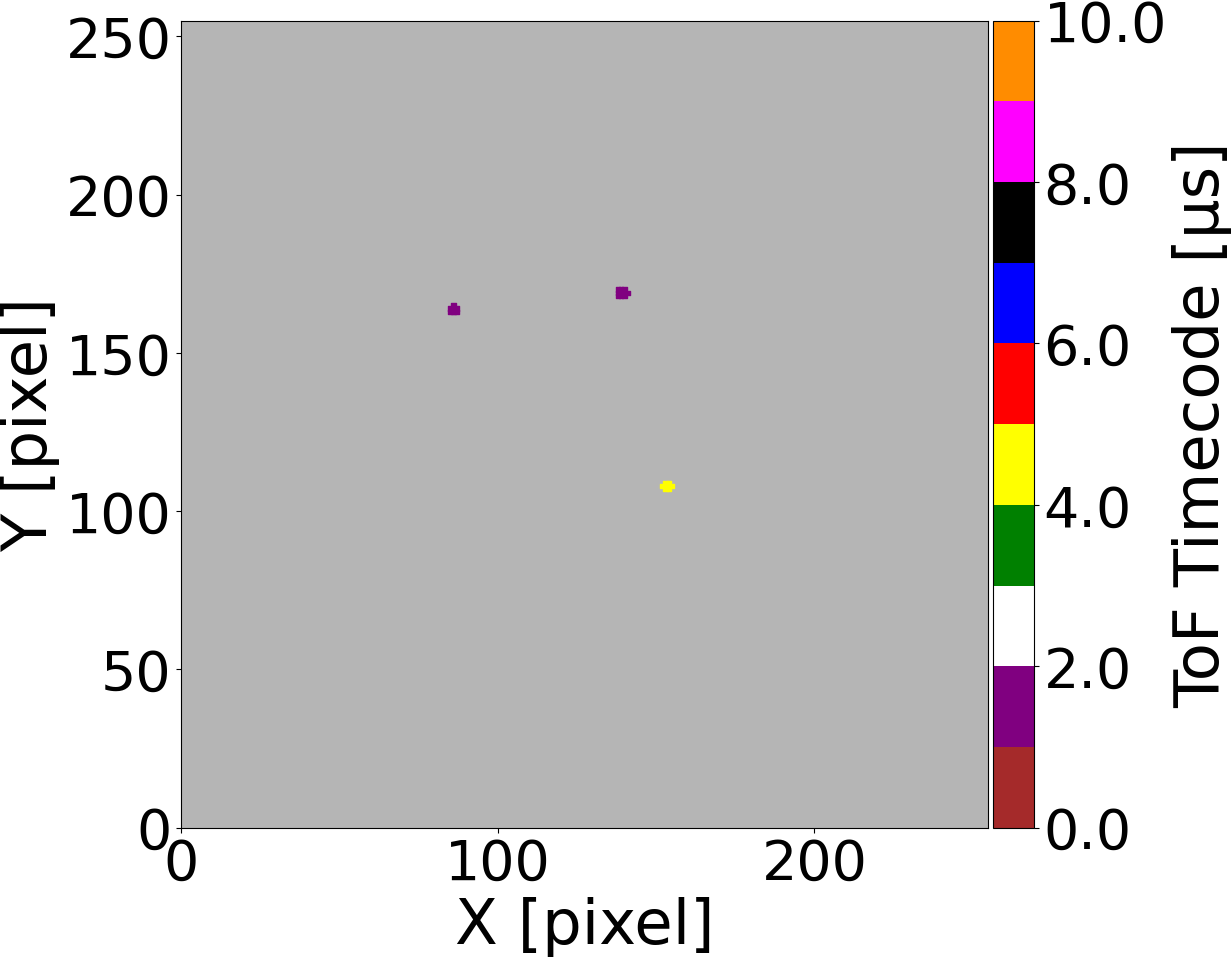}
\end{subfigure}\hspace{1mm}
\begin{subfigure}[b]{0.32\textwidth}
    \includegraphics[width=\textwidth]{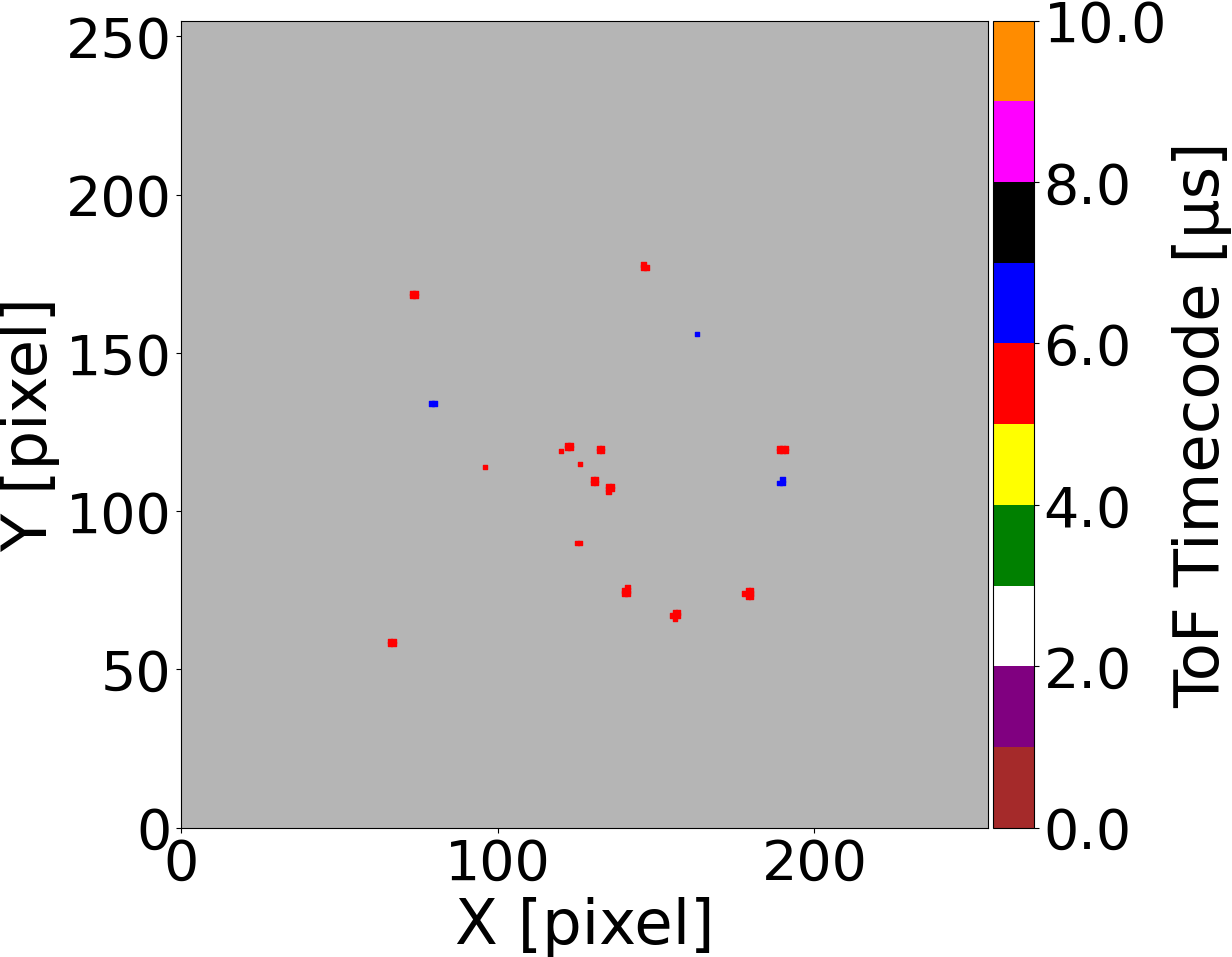}
\end{subfigure}
\par\smallskip
\text{(i) Examples of low-occupancy events, with up to $10^2$ pixels, originating from small nanoparticles.}
\smallskip

\begin{subfigure}[b]{0.32\textwidth}
    \includegraphics[width=\textwidth]{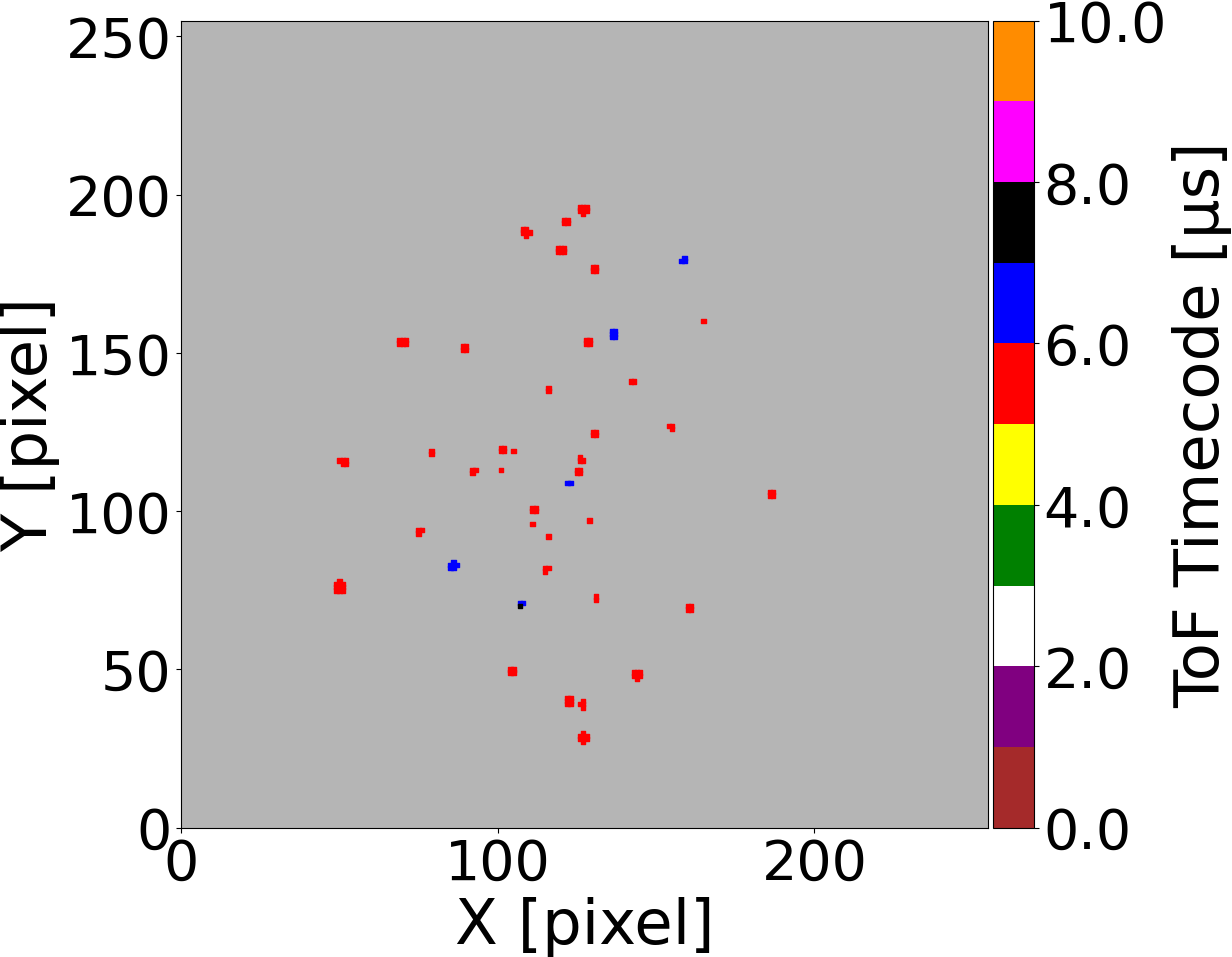}
\end{subfigure}\hspace{1mm}
\begin{subfigure}[b]{0.32\textwidth}
    \includegraphics[width=\textwidth]{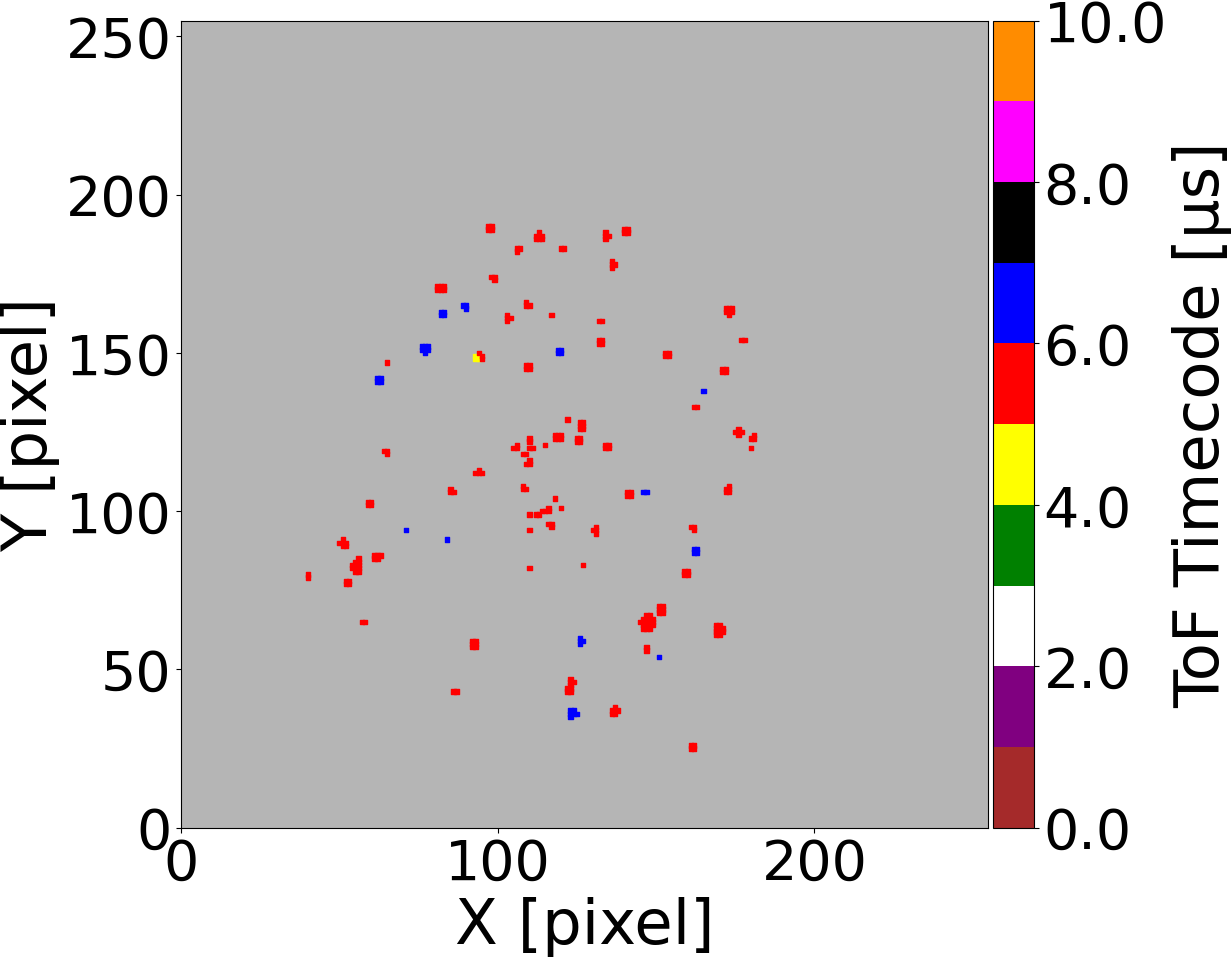}
\end{subfigure}\hspace{1mm}
\begin{subfigure}[b]{0.32\textwidth}
    \includegraphics[width=\textwidth]{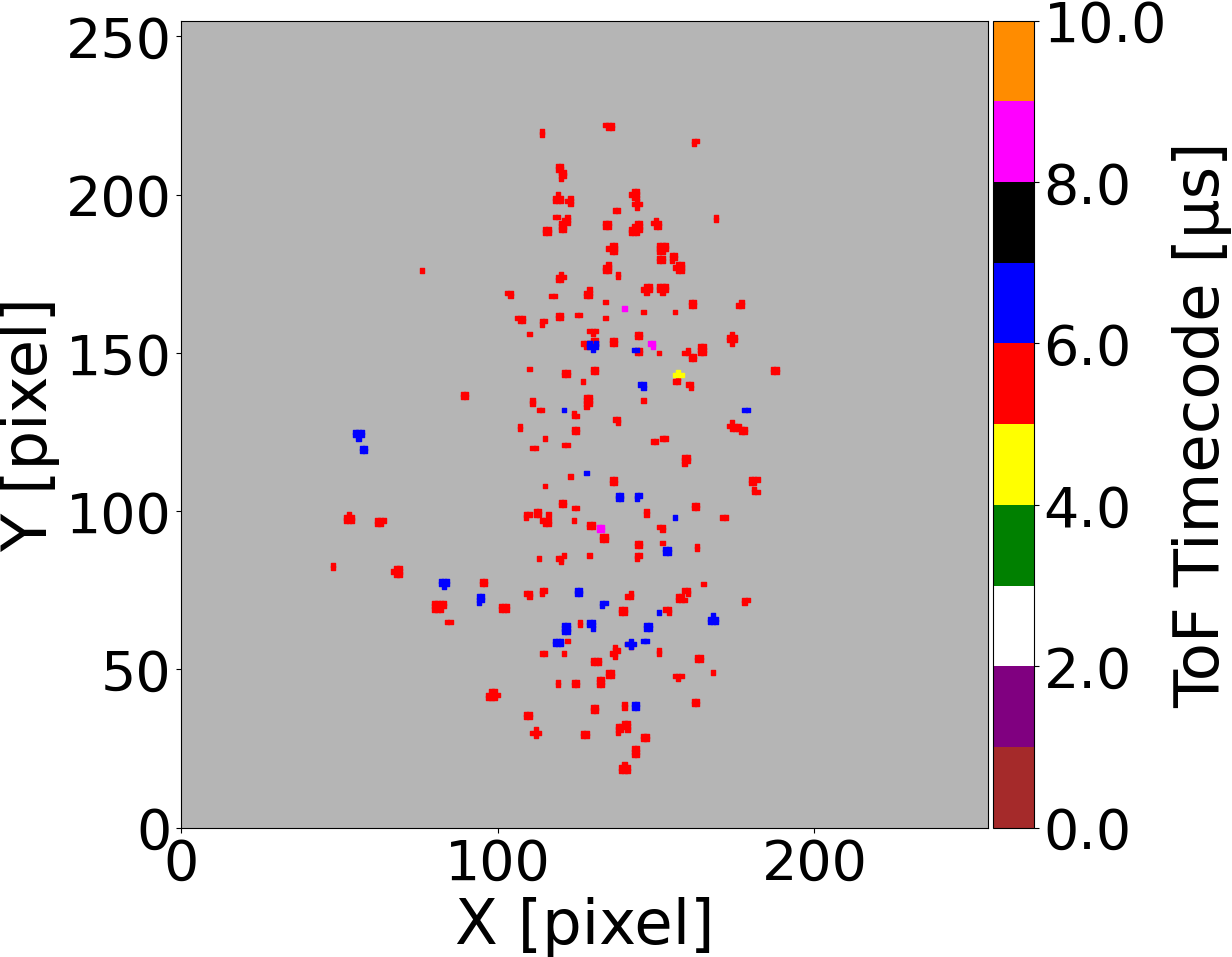}
\end{subfigure}
\par\smallskip
\text{(ii) Examples of events, with up to $10^3$ pixels, originating from medium size nanoparticles.}
\smallskip

\begin{subfigure}[b]{0.32\textwidth}
    \includegraphics[width=\textwidth]{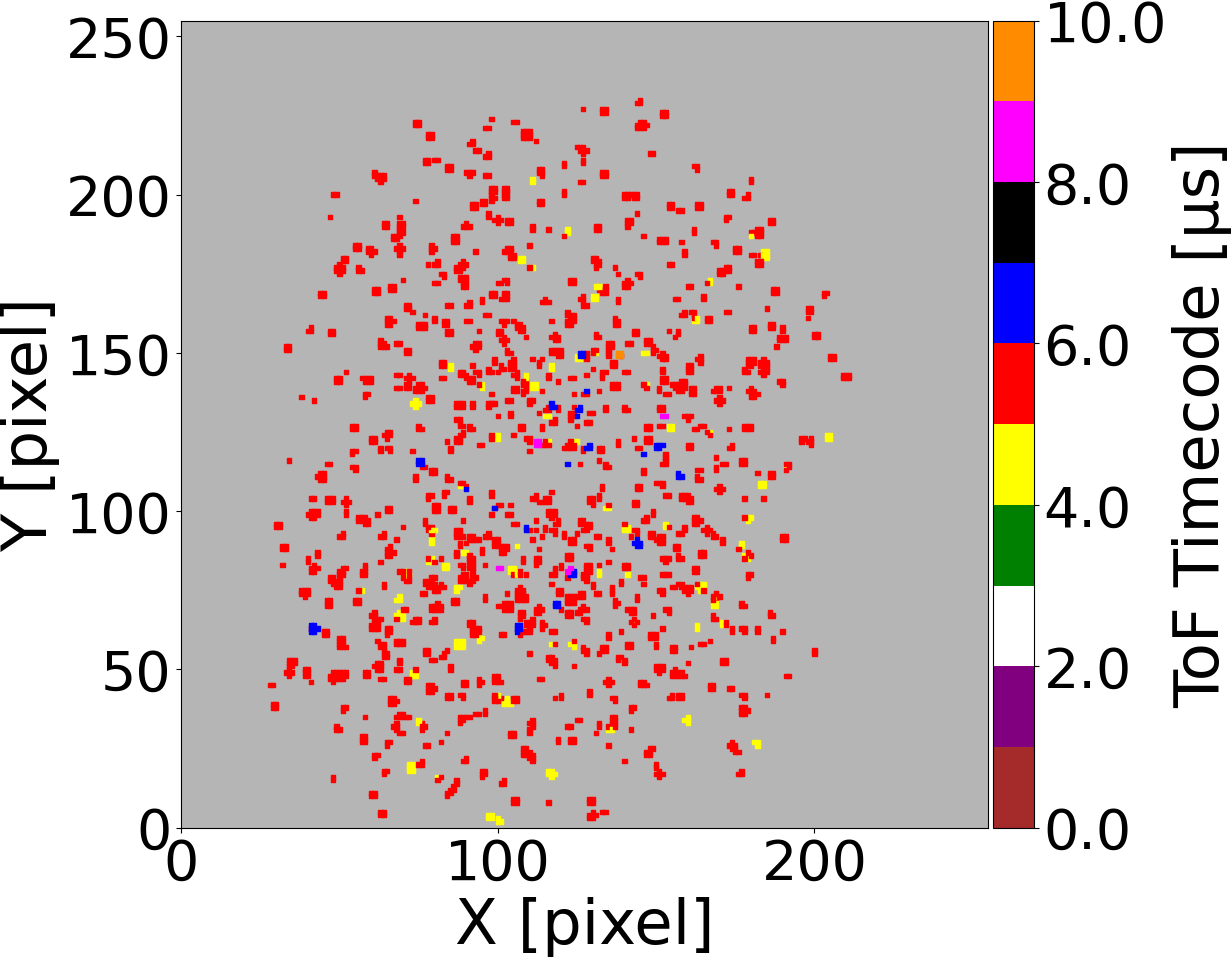}
\end{subfigure}\hspace{1mm}
\begin{subfigure}[b]{0.32\textwidth}
    \includegraphics[width=\textwidth]{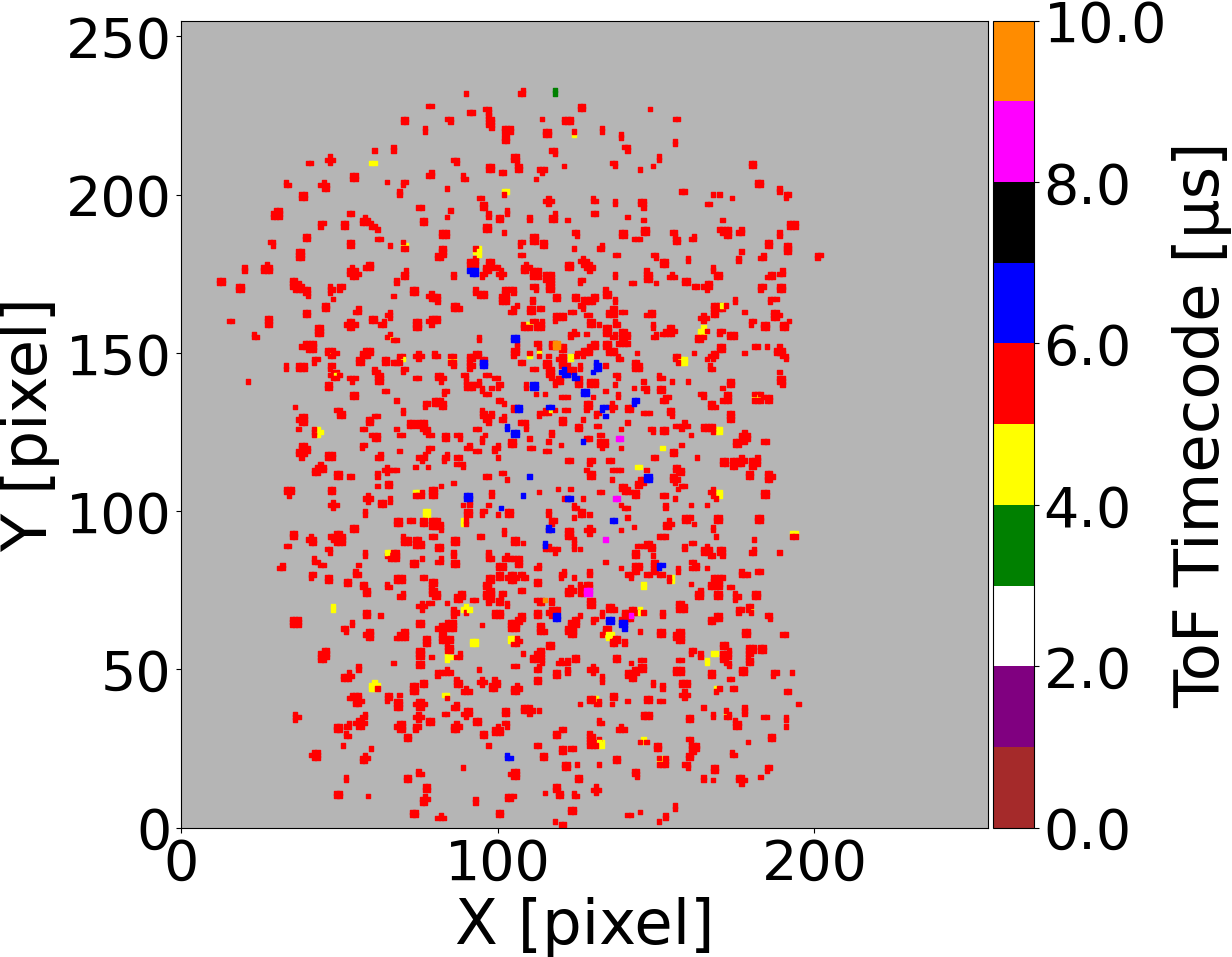}
\end{subfigure}\hspace{1mm}
\begin{subfigure}[b]{0.32\textwidth}
    \includegraphics[width=\textwidth]{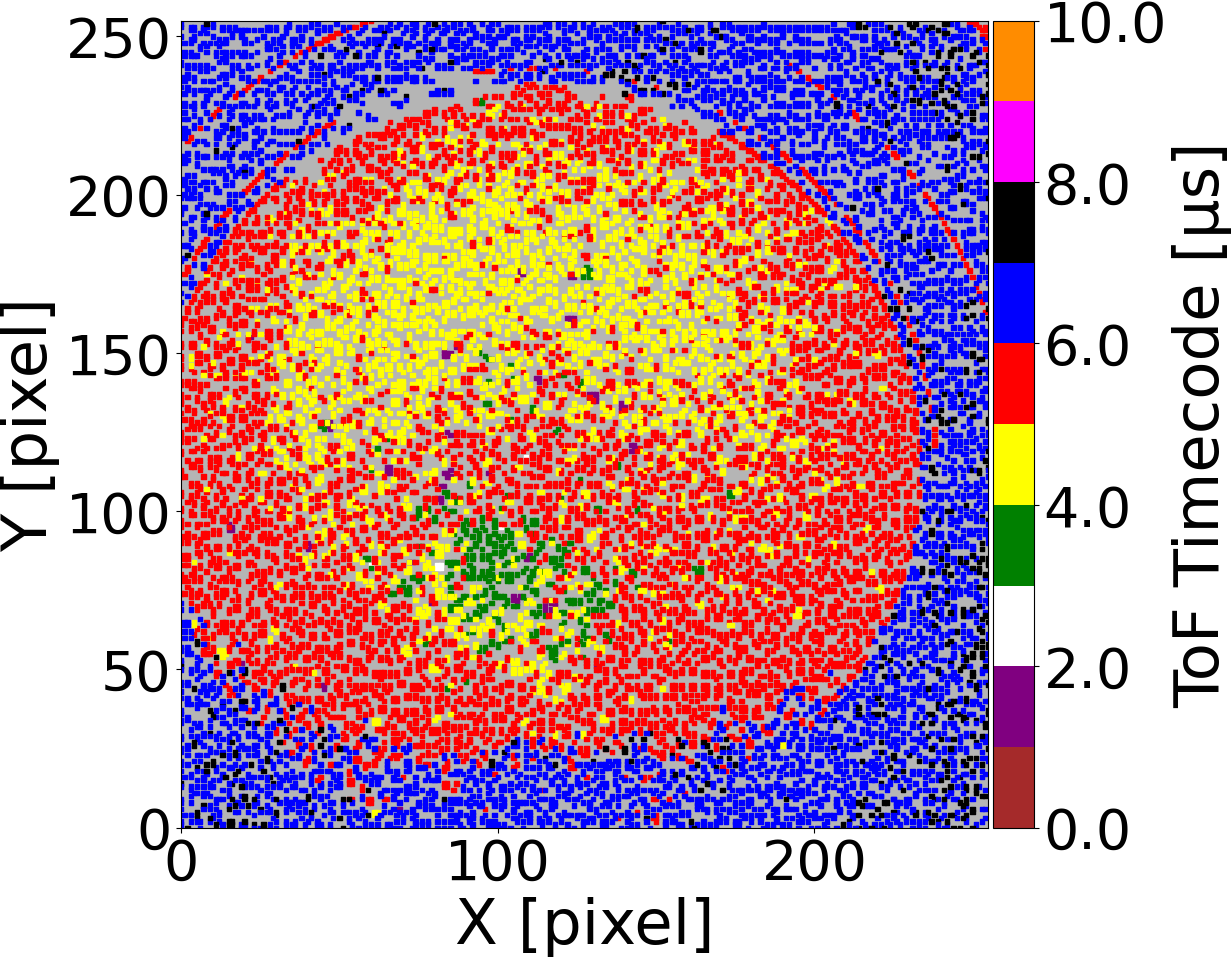}
\end{subfigure}
\par\smallskip
\text{(iii) Examples of high-occupancy events, with up to $10^5$ pixels, originating from large nanoparticles.}

\caption{Scatter plots of hits from individual triggers on the detector, illustrating the spatial distribution of ions across the camera for events of varying occupancy, represented by the colour scheme corresponding to ToF. The visible clusters vary in size and shape, from single-pixel hits to larger structures appearing due to the blending. \label{fig:varying occupancy}}
\end{figure}

As illustrated in Figure \ref{fig:sketch}, the recorded events originate from interactions of both small and large nanoparticles with the laser. The developed data analysis framework enables selection of these events and detailed examination of their spatial distributions across the detector. Examples of events ranging from near-zero to extremely large occupancies are shown in Figure \ref{fig:varying occupancy}, demonstrating the detector’s flexibility and robust capture capabilities. At the highest occupancies, the hit pixels, which belong to different ions, may start to blend (or merge) together. In principle, the blending can be mitigated by considering the timing and shape of individual clusters (of course, within limits) as discussed in \cite{fisher-levine2018}. Analysing these clusters with appropriate deblending algorithms, currently under development, should enable reconstruction of the underlying ion dynamics and behaviour. This also allows to define observables based on event multiplicity, their symmetries and other event properties.

\FloatBarrier

\begin{figure}[htbp]
\centering
\begin{subfigure}[b]{0.49\textwidth}
    \centering
    \includegraphics[width=\textwidth]{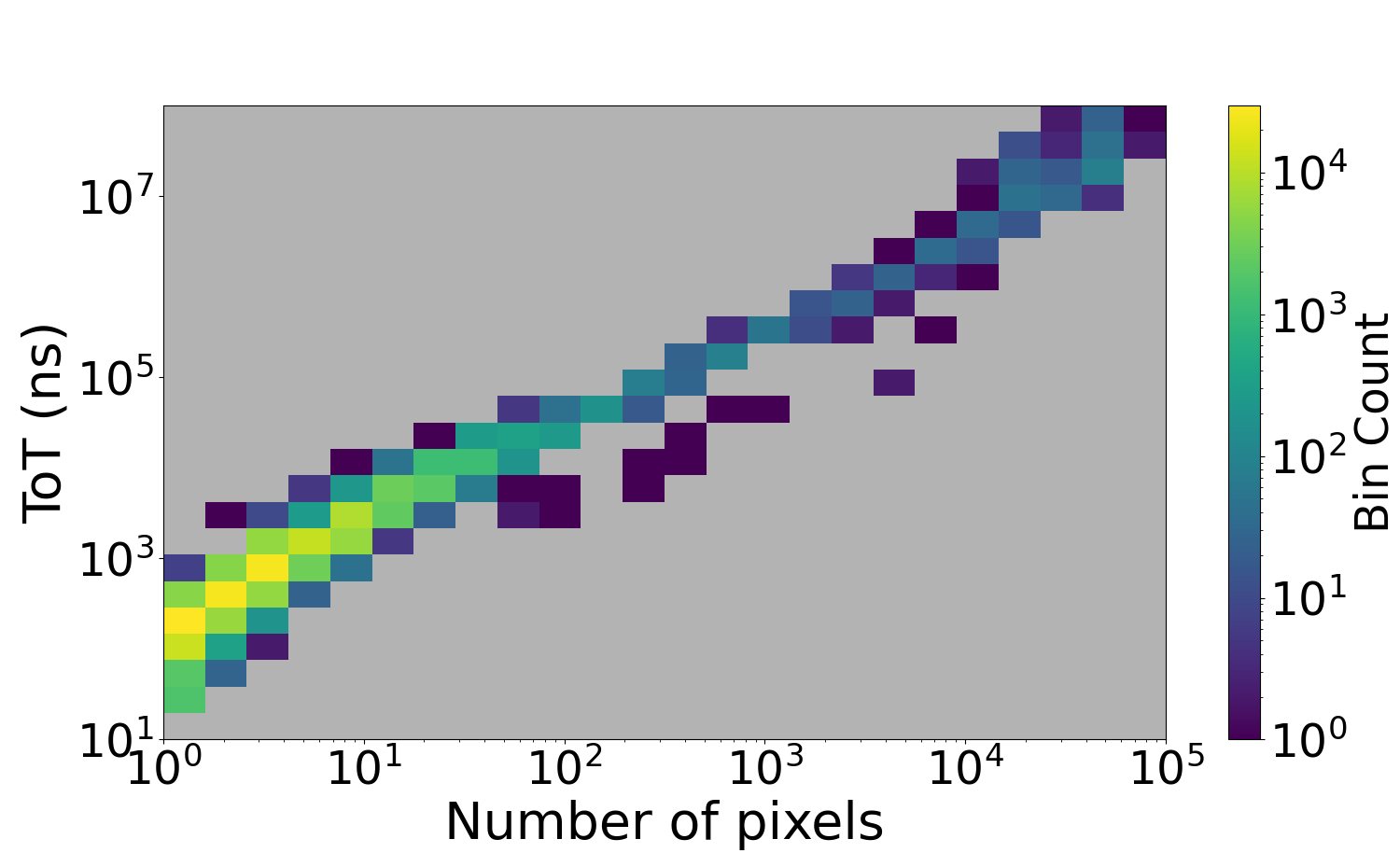}
\end{subfigure}
\hfill
\begin{subfigure}[b]{0.49\textwidth}
    \centering
    \includegraphics[width=\textwidth]{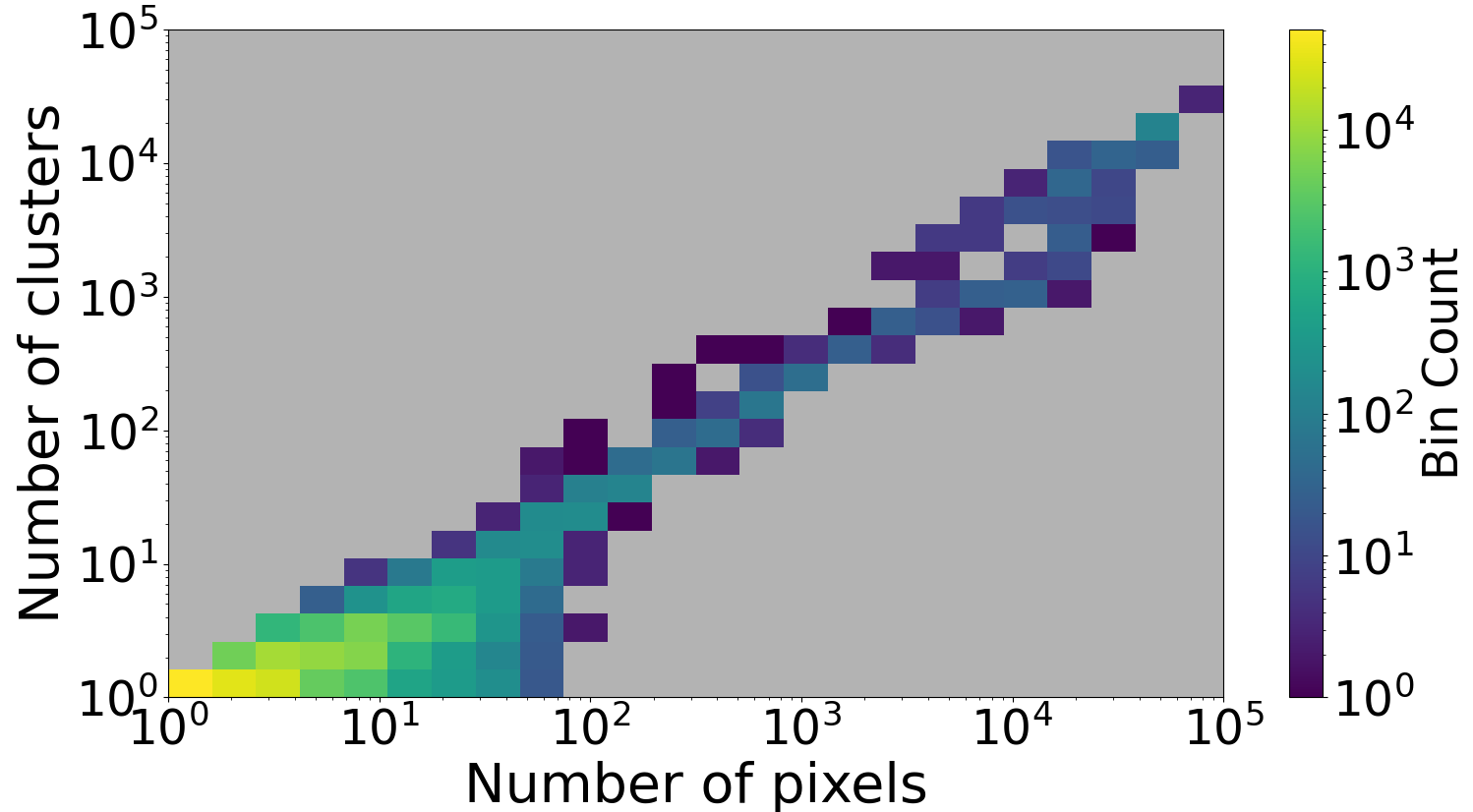}
\end{subfigure}

\caption{Correlation analysis of data corresponding to Kr$^+$ mass peak. Left: total event ToT (so the total produced light intensity in the ion detector) as a function of the total number of pixels hit in the event. Right: Number of clusters in an event as a function of total number of pixels hit in the event. The distributions reveal strong correlations between the observables and also the presence of events with high occupancy of krypton ions, which can be easily identified and separated from events with a small number of hit pixels.
}
\label{fig:possibilites}
\end{figure}

After the data set is corrected for the time-stamp shifts, a wide range of further analysis becomes available, such as selection of specific ion species, investigating clusters of different sizes, structures, and distributions, and exploring potential correlations. Figure \ref{fig:possibilites} illustrates two examples of such analysis where we used hits only from the Kr$^+$ mass peak with the corresponding mass range defined as in Figure~\ref{fig:ions}. The left graph shows the total event ToT (so the total  light intensity produced in the ion detector) as a function of the total number of pixels hit in the event. The right graph shows the number of reconstructed clusters (which can be interpreted as the total number of ions in the event) as a function of the total number of pixels hit in the event. The distributions reveal strong correlations between the observables and also the presence of events with high occupancy of krypton ions, which can be easily identified  and separated from events with a small number of hit pixels. Further analysis of their properties is in progress.

\section{Conclusion}
\label{sec:conclusion}

We have successfully demonstrated the capabilities of the Timepix3 optical camera to investigate ultrafast ion dynamics in substrate-free nanoparticles under extreme occupancy conditions. The detector captured events ranging from near-zero to extremely high occupancies, approaching the readout limits. 
The Tpx3Cam demonstrated its robustness to handle extreme occupancy variations in nanoparticle-laser interaction experiments. The detector's performance near its readout limits revealed both capabilities and constraints, which are crucial to understand for the design of future experiments. The applied correction algorithm successfully restored data that would otherwise be misinterpreted, ensuring data integrity across the entire experimental range.

The ability to distinguish between small and large nanoparticle interactions within the same experimental run provides control and enables systematic studies of size-dependent ion production mechanisms. The stochastic nature of nanoparticle-laser interactions, with variable signal levels, creates ideal conditions to demonstrate the advantages of time-stamping detectors. The successful handling of such statistically varying data validates the detector's suitability for complex experimental scenarios.


\acknowledgments

The authors thank the staff of the ELI Beamlines Facility, a European user facility operated by the Extreme Light Infrastructure ERIC, for their support and assistance. This research was supported by the Czech Science Foundation (GACR) under Project No. 25-15534M and Czech Ministry of Education, Youth and Sports Project No. LM2023040 CERN-CZ. We thank Lou-Ann Pestana De Sousa for help with graphics.


\bibliographystyle{JHEP}
\bibliography{biblio.bib}






\end{document}